\documentclass[10pt,letterpaper,twocolumn]{article} 

\usepackage{ol2}
\usepackage{amsmath}
\usepackage{hyperref}
\begin{document}

\twocolumn[ 

\title{Optical Rotation Quasi-Phase-Matching for Circularly Polarized High Harmonic Generation}


\author{Lewis Z. Liu,$^{*}$ Kevin O'Keeffe, and Simon M. Hooker}

\address{
Clarendon Laboratory, University of Oxford Physics Department,\\
 Parks Road, Oxford OX1 3PU, United Kingdom  \\
$^*$Corresponding author: L.Liu1@physics.ox.ac.uk }

\begin{abstract}The first scheme for quasi-phase-matching high harmonic generation of circularly
polarized radiation is proposed: optical rotation
quasi-phase-matching (ORQPM). In ORQPM propagation of the driving
radiation in a system exhibiting circular birefringence causes its
plane of polarization to rotate; by appropriately matching the
period of rotation to the coherence length it is possible to avoid
destructive interference of the generated radiation. It is shown
that ORQPM is approximately 5 times more efficient than conventional
QPM, and half as efficient as true phase-matching. 

This paper was published in Optics Letters and is made available as an electronic reprint with the permission of OSA. The paper can be found at the following URL on the OSA website: \url{http://www.opticsinfobase.org/ol/abstract.cfm?uri=ol-37-12-2415}. Systematic or multiple reproduction or distribution to multiple locations via electronic or other means is prohibited and is subject to penalties under law.  Please cite as L. Z. Liu, K O'Keeffe, S. M. Hooker, Optics Letters, Vol. 37, Issue 12, pp. 2415-2417 (2012)\end{abstract}

\ocis{190.2620, 340.7480.}

 ] 

\noindent When an intense laser pulse on the order of $10^{14}$ W/cm$^2$ is focussed into a low density
gas high order harmonics of the fundamental driving field can be
produced.  This high harmonic generation results in coherent
radiation extending to soft x-ray wavelengths.  As such it is an
attractive source of EUV radiation and has applications including
time-resolved science \cite{Cavalieri}, ultrafast holography
\cite{RaHolography}, and coherent diffractive imaging
\cite{Sandberg}.

However, without additional techniques, HHG is very inefficient;
typical photon conversion efficiencies are $10^{-6}$ for generated
photons with energies of order $100$ eV, decreasing to $10^{-15}$
for generation of 1 keV radiation. This inefficiency is largely due
to the fact that the driving and harmonic fields have different
phase velocities; as such they develop a phase mismatch, causing the
intensity of the generated harmonics to oscillate with propagation
distance with a period $2 L_c = 2 \pi / \Delta k$.  The wave vector
mismatch $\Delta k$ is in general non-zero owing to dispersion in
the target material, geometric dispersion, and waveguide dispersion;
it is given by $\Delta k = k(q \omega) - q k(\omega)$, where
$k(\omega)$ is the wave vector of radiation of angular frequency
$\omega$ and $q$ is the harmonic order.

The efficiency of HHG can be greatly increased by true
phase-matching, i.e. balancing dispersion in the system so that
$\Delta k = 0$. In the case of harmonics generated in a hollow core waveguide true phase-matching may be achieved by tuning the pressure in the waveguide, enabling the quadratic growth of the harmonic intensity with propagation distance. 
\cite{Murnane1998}. However, true phase-matching may only be achieved up
to a critical ionization level, above which it is no longer possible
to achieve $\Delta k=0$, placing a limit on the maximum harmonic
order which can be phase-matched.  An alternative approach to improve the efficiency of HHG is to employ the technique of quasi-phase-matching, although QPM is not as efficient as
true phase-matching. In QPM, HHG is suppressed in regions where the
locally generated harmonic is out of phase with the harmonic beam.
By suppressing HHG in multiple out-of-phase regions the harmonic
intensity grows monotonically with $z$. Techniques for QPM include
the use of counter-propagating pulses \cite{Zhang}, multi-mode
beating \cite{Zepf2007}, modulated waveguides \cite{ModWaveguide},
and modulated gas density \cite{Seres}, and static electric fields \cite{Serrat}.

Recently we proposed a new QPM technique: polarization beating QPM
(PBQPM) \cite{Patent, LiuPRAPBQPM}. In this approach, a linear
birefringent system modulates the polarization of the driving pulse,
causing it to beat between linear and elliptical. Because harmonic
generation is suppressed for elliptically polarized light, QPM can
be achieved if the period of the polarization beating is suitably
matched to the coherence length.  In this paper, we propose a novel,
more efficient, QPM scheme that enables the generation of circularly
polarized high harmonics: Optical Rotation Quasi Phase Matching
(ORQPM) \cite{PatentORQPM}.  ORQPM utilizes a waveguide with circular
birefringence (as opposed to linear birefringence), which causes the plane of polarization of linearly
polarized light to rotate with propagation distance at a constant
rate, with period $2 L_r$. By matching $L_r = L_c$, the generated
harmonics will grow monotonically. As we show in detail below, ORQPM
allows harmonics beyond the true phase-match limit to be generated with
comparable efficiencies to that obtained with true phase-matching.
Moreover, ORQPM is the first QPM scheme to generate circularly
polarized high harmonics; bright sources of circularly polarized
soft x-radiation would find widespread application in studies of
ultrafast spin dynamics \cite{Koopmans} and nano-lithography
\cite{PetraLithography}.  We also note that a
polarization gating technique in metallic macrostructures has been
proposed to generate circuarly polarized high harmonics
\cite{HusakouPolarizationGating} and that second harmonic generation
quasi-phase matching has been previously achieved in an optically
active solid media \cite{BussonChiral}.  In this paper we describe ORQPM in detail,
demonstrate its operation by means of a simple model, and discuss
three techniques by which it might be realized.

\begin{figure}[htb]
\centerline{\includegraphics[width=6.5cm]{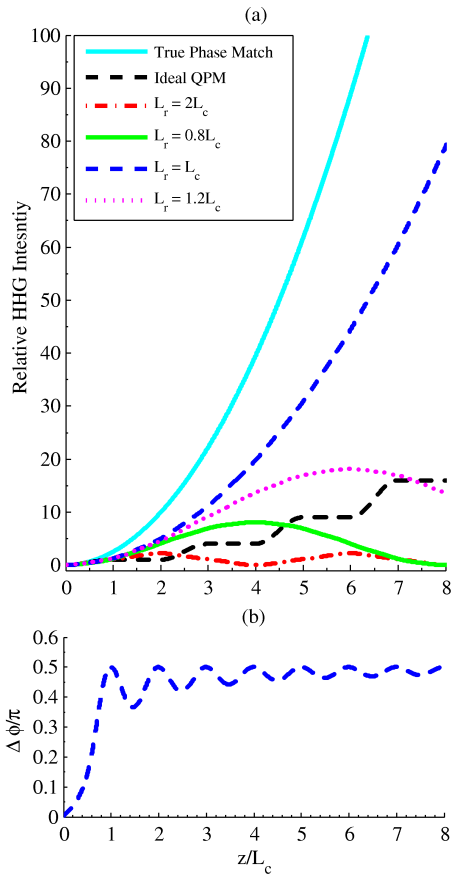}}
\caption{(a) Relative HHG intensity for true phase matching (solid
cyan line), ideal QPM (dashed black line), and ORQPM for $L_r = 2
L_c$ (dot-dashed red line), $0.8 L_c$ (solid green line), $L_c$
(dashed blue line), $1.2 L_c$ (dotted pink line) normalized to the
the output after a single coherence length with no phase matching.
(b) Phase difference $\Delta \phi = \arg(\hat{\xi_x}) -
\arg(\hat{\xi_y})$ between the $x$ and $y$ polarization components
as a function of $z$ for $L_r = L_c$. \label{fig:GainAndDeltaPhi}}
\end{figure}

The driving electric field can be written as a linearly polarized
wave with a plane of polarization that rotates with propagation
along the ${z}$ axes with period $ L_r =  \pi / \nu$, where $\nu$ is the optical rotation power:
\begin{equation} \vec{E}_{\mathrm{driver}}(z, t) = E_0 e^{i(\beta z - \omega t)} \left( \begin{array}{ll}
\cos(\nu z) \\ \sin(\nu z) \end{array} \right). \label{eqn:driver}
\end{equation} where $\beta = k(\omega)$ is the propagation constant and $E_0$ is the field amplitude.  For the
purpose of demonstrating ORQPM, we will assume that the damping of
the driving radiation is negligible.

In the slowly-varying envelope approximation, the electric field of
the $q$th harmonic can be written as:
\begin{equation} \vec{E}_q (z, t) = \vec{\xi}(z, t) e^{i [k(q \omega)z
- q\omega t]} \end{equation} where $\vec{\xi}$ is the electric field
envelope. The Rayleigh range of the harmonic beam will be
approximately $q$ times that of the driving beam; as such we will
ignore the effect of the waveguide on the wave-vector $k(qw)$ of the
harmonic beam. We note that in cases where $k(qw)$ is affected by
the waveguide, ORQPM will still occur if there is a difference
between the circular birefringence experienced by the driving and
generated beams.

Under a continuous-wave approximation, the amplitude of the ${x}$-
and ${y}$- components of the harmonic envelope will grow according
to,
\begin{equation}\left\{
\begin{array}{ll}
    \frac{\partial\xi_x}{\partial z} = A \Lambda_x(z) e^{-i \Delta k z}
    \\
    \frac{\partial\xi_y}{\partial z} = A \Lambda_y(z) e^{-i \Delta k z} \end{array} \right
    . \label{eqn:diffeq}
    \end{equation}
where $A$ is a normalization constant, and $\Lambda_x$ and
$\Lambda_y$ are the relative source terms.  In this case the driving
radiation has the same intensity at all points along the system, and
is always linearly polarized. The locally generated harmonics will
have a polarization parallel to that of the driving beam, and hence
$\Lambda_x = \cos (\nu z)$ and $\Lambda_y =\sin (\nu z)$.

Assuming that $A$ and $\Delta k$ are constant, Eqn.
\eqref{eqn:diffeq} yields,
\begin{equation}\left\{ \begin{array}{ll}
    \xi_x(z)= i A \frac{\Delta k - e^{i \Delta k z} [\Delta k
    \cos(\nu z) + i \nu \sin(\nu z) ]}{\Delta k^2 - \nu^2} \\
    \xi_y(z) = A \frac{-\nu + e^{i \Delta k z} [\nu \cos(\nu z) + i
    \Delta k \sin(\nu z)]} {\Delta k^2 - \nu^2} \end{array} \right.
    .
    \label{eqn:envsolve} \end{equation}
ORQPM corresponds to setting $\nu \rightarrow \Delta k$, whereupon,
\begin{equation}\left\{ \begin{array}{ll}
    \hat{\xi}_x(z) = \lim_{\nu \to \Delta k}
     \left[\xi_x(z)\right] = A \frac{ 2 \Delta k z- i +
    i e^{-2 i \Delta k z}}{ 4 \Delta k}  \\ \\
    \hat{\xi}_y(z) = \lim_{\nu \to \Delta k}
     \left[\xi_y(z)\right] = A
    \frac{-2 i \Delta k z+1-e^{-2 i \Delta k z}}{4 \Delta k}
    \end{array} \right. .
    \label{eqn:envxihat} \end{equation}
    From this the intensity, $\hat{I}(z)  =
    \hat{\xi_x}\hat{\xi_x}^*+\hat{\xi_y}\hat{\xi_y}^*$ is found to be:
\begin{equation}    \hat{I}(z) = A^2\left[\frac{1}{2}z^2 + \frac{1 - \cos(2 \Delta k z)}{4 (\Delta
    k)^2}\right] .\end{equation}

We see that the growth of the harmonic intensity comprises a
quadratic term, which dominates at large $z$, plus a weak
co-sinusoidal modulation. It is clear that in the limit of large
$z$, ORQPM is half as efficient as would be true phase-matching ---
i.e. setting $\Delta k = 0$ --- under otherwise identical
conditions.

These analytical results are confirmed by the results of numerical
integration of Eqn. \eqref{eqn:diffeq}, as shown in Fig.
\ref{fig:GainAndDeltaPhi}. When properly matched, ORQPM causes the
intensity of the harmonic to grow almost monotonically, and it may be seen that at large $z$ that the intensity is half that which would be obtained with
true phase-matching. Notice that with ORQPM the harmonic intensity
grows $\pi^2 / 2 \approx 5$ times faster than ideal QPM, defined to
be the square-wave modulation of the local harmonic generation with
a period $2L_c$ and complete suppression of harmonic generation in
the out of phase zones.

In Fig. \ref{fig:GainAndDeltaPhi}(b) the phase difference between
the two polarization states of the harmonic is illustrated; it is
seen that after a few coherence lengths this phase difference is
close to $\pi/2$, corresponding to circular polarization of the
harmonic. This can also be seen in Eqn \eqref{eqn:envxihat}, where
the dominating terms for the $x-$ and $y-$ components of the
envelope function, $\frac{A}{2}z$ and $-\frac{iA}{2}z$ respectively,
have the same amplitude but are $\pi/2$ out of phase, corresponding
to circular polarization. For fixed $t$, the rotation with $z$ of
the plane of polarization of the harmonic radiation is in the same
sense as that of the driving beam, i.e. right(left)-hand polarized
harmonics are generated by right(left)-rotating driving radiation.

In order to achieve ORQPM in practise it would be necessary to guide a
laser pulse over an extended region using a waveguide with circular
birefringence that has a rotation length shorter than $~1$mm.  Such a
waveguide could be constructed in a number of ways. One possibility
for achieving ORQPM is a waveguide constructed from a material with
a high Verdet constant $V$, which would allow ORQPM to be achieved
by tuning the applied magnetic flux density $B$ until $\nu = V B =
\Delta k$. We note that Verdet constants lie in the range $100$ deg
mm$^{-1}$ T$^{-1}$ to $10^5$ deg mm$^{-1}$ T$^{-1}$
\cite{KoeckelberghsFaraday}. Solid-core waveguides exhibiting
Faraday rotation have been developed,although the hollow-core systems which would be required for ORQPM have yet to be developed \cite{ZamanFaraday, YuFaraday}.

Alternatively, the waveguide walls could be constructed from optically
active materials, which have rotary powers in the range $10^{2}$ to
$10^4$ deg mm$^{-1}$ \cite{JerphagnonChiral, KoshimaChiral}.
Further, we note that polarization rotating photonic crystal fibres
have been developed with $L_r<1$mm
\cite{HameedChiral} and that HHG has been acheived using PCFs \cite{Heckl}.

Finally, we also note that ORQPM could be achieved in
non-birefringent waveguides by exciting two circularly polarized
waveguide modes with different mode velocities. The resultant
superposition of these two modes will be linearly polarized with 
rotating polarization.

As for any QPM scheme limits will be set by absorption of the
harmonics and variation of the coherence length \cite{OKeeffePTQPM}. As for other
QPM schemes based on polarization-control of the driving laser, the
distance over which ORQPM can be achieved may be limited by relative
slippage of the constituent polarizations.  However, we note that in
principle this limit can be avoided by reversing the rotation of the
driving field when the two modes have slipped apart, although the
polarization of the resultant harmonic field may no longer be
perfectly circularly polarized.  Finally, we point out that ORQPM may also be limited by chromatic despersion from either waveguide or Faraday effects. 

In this paper, we have proposed a novel QPM scheme for high harmonic
generation that relies on rotation of the driving radiation using a
circularly birefringent waveguide. By matching the rotation length
to the coherence length, circularly polarized harmonics may be
produced with an efficiency up to 50\% of that obtained with true
phase-matching and 5 times more efficient than conventional QPM.
These findings were confirmed by numerical simulations.  Three
systems in which ORQPM might be achieved have been identified;
exploration of these will form the basis of future work.

The authors would like to thank EPSRC for support through grant No.
EP/GO67694/1 and Merton College, Oxford for financial support.

\bibliographystyle{osajnl}


\pagebreak

\section*{Informational Fourth Page}
In this section, please provide full versions of citations to
assist reviewers and editors (OL publishes a short form of
citations) or any other information that would aid the peer-review
process.

\end{document}